\documentstyle[11pt,twoside,fleqn,unicrc,amsmath,bbm]{article}
\author{H. Reinhardt\\Institute for Theoretical Physics\\T\"ubingen
  University\\Auf der Morgenstelle 14\\D-72076 T\"ubingen, Germany}
\title{Magnetic Monopoles and the Topology of Gauge Fields\thanks{Talk
    presented at the Int. Conf. QCD 98, Montpellier, July 1998.}\thanks{Work supported
    DFG (Re 856/1--3).}}
\begin{document}
\begin{abstract}
Lattice calculations performed in Abelian gauges give strong evidence that
confinement is realized as a dual Meissner effect, implying that the Yang--Mills
vacuum consists of a condensate of magnetic monopoles. We show in Polyakov gauge 
how the Pontrjagin index of the gauge field is related to the 
magnetic monopole charges.
\end{abstract}
\maketitle

\section{Introduction}
There are two basic low energy features of strong interactions, which should be
explained by QCD: confinement and spontaneous breaking of chiral symmetry.
While the latter can be more or less explained in terms of instantons \cite{1},
confinement is much less understood. Recently, however, much evidence has been
accumulated for the idea that confinement is realized as a dual Meissner effect
\cite{2}, at least in the so-called Abelian gauges \cite{3}. In these gauges
magnetic monopoles arise, which have to condense in order to form the dual
superconducting ground state.

From the conceptional point of view it is unsatisfactory that spontaneous
breaking of chiral symmetry and confinement is attributed to different types of
field configurations. This is because lattice calculations show that both
phenomena occur at about the same energy scale. In fact, 
lattice calculations indicate that the deconfinement phase transition and the
restoration of chiral symmetry occur at about the same temperature. 
Furthermore, for the spontaneous breaking of chiral symmetry the topological
properties of the instantons are crucial, giving rise to zero modes of the
quarks, which in turn lead to a non-zero level density at zero virtuality. The 
latter is sufficient to ensure a non-zero quark condensate, which is the order
parameter of chiral symmetry breaking \cite{1}. On the other hand, magnetic 
monopoles are long-range fields, which should also be relevant for the global
topological properties of gauge fields. We can therefore expect an intimate
relation between magnetic monopoles and the topology of gauge fields. It is the aim
of my talk to explain how the non-trivial topology of gauge fields is generated
in Abelian gauges by magnetic monopoles \cite{4}.

\section{Emergence of Magnetic Monopoles in Abelian Gauges}
Magnetic monopoles arise in the so-called Abelian gauges, which fix the coset
$G/H$ of the gauge group $G$, but leave the Abelian gauge invariance with
respect to the Cartan subgroup $H$ \cite{3}. Recent lattice calculations show 
that the dual Meissner effect is equally well realized in all Abelian gauges
studied \cite{5}, while the Abelian and monopole dominance is more pronounced in the
so-called maximum Abelian gauge \cite{2}. Here, for simplicity, we will use the 
Polyakov gauge, defined by a diagonalization of the Polyakov loop
\begin{eqnarray}
\Omega (\vec x) &=& P \exp \left( - \int^T_0 dx_0 A_0 \right)\\
 &=& V^\dagger \omega V
\to \omega, \label{X1}
\end{eqnarray}
where $\omega \in H$ is the diagonal Polyakov loop and the matrix $V
\in G/H$ is obviously defined only up to an Abelian gauge transformation $V
\to gV,g\in H$. This gauge is equivalent to the condition
\begin{eqnarray}
A^{\text{ch}}_0 = 0, \quad \partial_0 A^{\text{n}}_0 = 0,
\end{eqnarray}
where $A^{\text{n}}_0$ and $A^{\text{ch}}_0$ denote the diagonal (neutral with respect to $H$)
and off-diagonal (charged) parts of the gauge field, respectively.

When the Polyakov loop (\ref{X1}) becomes a center element of the gauge group
at some isolated point $\bar x$ in $3$-space there is a topological
obstruction in the diagonalization and the induced gauge field
\begin{eqnarray}
{\cal A} = V \partial V^\dagger, \label{X4}
\end{eqnarray}
arising from the gauge transformation $V \in G/H$ which makes the Polyakov
loop diagonal develops a magnetic monopole \cite{3,4}.

Let us illustrate this for the gauge group $G=SU(2)$, where the Cartan subgroup
is $H = U(1)$ and the Polyakov loop can be parametrized as 
 \begin{eqnarray}
\Omega (\vec x) &=& \exp (i \vec \chi (\vec x) \vec \tau)\nonumber\\
 &=& \cos \chi + i \hat \chi
\vec \tau \sin \chi\nonumber\\
(\chi &=& \vert \vec \chi \vert, \hat \chi = \vec \chi / \chi),
\end{eqnarray}
with $\tau_a, a = 1, 2, 3$ being the Pauli matrices. In this case the matrix $V
\in G/H$ can be chosen as \cite{4}
\begin{eqnarray}
V (\vec x) &=& e^{i \Theta \tau_{\phi}/2}\\
(\tau_\phi &=& \sin \phi \tau_1+ \cos \phi \tau_2)\nonumber,
\end{eqnarray}
where $\Theta$ and $\phi$ denote the polar and azimuthal angles of
the color unit vector $\hat \chi$, respectively, and the induced gauge field
(\ref{X4}) acquires the form
\begin{eqnarray}
{\cal A}_\mu &=& - \frac{i}{2} \tau_\phi \partial_\mu \Theta + \frac{i}{2} \sin
\Theta \tau_\rho \partial_\mu \phi\nonumber\\
 &+& \frac{i}{2} (1- \cos \Theta) \partial_\mu
\phi \tau_3\label{X6}\\
(\tau_\rho&=&-\partial_\phi\tau_\phi)\nonumber.
\end{eqnarray}
At those isolated points $\bar x_i$ in space where $\chi (\bar x_i) = \pi n_i$,
$n_i$ being an integer, and consequently the Polyakov loop becomes a center element
\begin{eqnarray}
\Omega (\bar x_i) = (-1)^{n_1},\label{X3}
\end{eqnarray}
the Abelian part of the induced gauge potential ${\cal A}$ (\ref{X6}) has the
form of a magnetic monopole.
To be more precise the Abelian part of the magnetic field, $\vec B = \vec \nabla
\times {\cal A}^3$ develops an ordinary Dirac monopole
with a Dirac string along the negative 3-axis. Note that the
$U(1)$ gauge freedom in the definition of $V \in G/H$ can be exploited to
move the Dirac string around. The Abelian component of the non-Abelian part of
the magnetic field $\vec b = (\vec {\cal A} \times \vec {\cal A})^3$,
contains an oppositely directed monopole field, but without a Dirac string. In
the Abelian component of the total magnetic field $B^3_i = \frac{1}{2}
\epsilon_{ijk} F_{jk} [{\cal A}]^3$ the monopole field cancels and only the Dirac
string survives \cite{6}. This is easily understood by noticing that the 
field ${\cal A}$ (\ref{X4}) is a pure gauge and its field strengths 
$F_{ij} [{\cal A}]$ should hence vanish unless there are singularities, which
here occur on the Dirac string.

An important property of the magnetic monopoles arising in the Abelian gauges
is that their magnetic charge
\begin{eqnarray}
m [V] &=& \frac{1}{4 \pi} \int_{\bar{S}_2} d \vec{\Sigma} \vec B^3\nonumber\nonumber\\
 &=& - \frac{1}
{8 \pi} \int_{S_2} d \Sigma_i \epsilon_{ijk} [{\cal A}_i, {\cal A}_j]\label{X10}
\end{eqnarray}
is topologically quantized and given by the winding number $m [V] \in \Pi_2
(SU(2)/U(1))$ of the mapping $V(\vec x) \in SU(2)/U(1)$ \mbox{\cite{7,4,6}}. In 
the above equation $S_2$ is an infinitesimal $2$-sphere around the monopole position and
$\bar S_2$ arises from $S_2$ by leaving out the position of the Dirac string.

\section{Magnetic Monopoles as Sources of Non-Trivial Topology}
Since the magnetic monopoles are long-range fields, we expect that they are
relevant for the topological properties of gauge fields. As is well known, the
gauge fields $A_\mu (x)$ are topologically classified by the Pontrjagin index
\begin{eqnarray}
\nu &=& - \frac{1}{16 \pi^2} \int d^4 x Tr F_{\mu \nu} \tilde F_{\mu \nu}\label{X11}\\
(\tilde F_{\mu \nu} &=& \frac{1}{2} \epsilon_{\mu \nu \kappa \lambda} F_{\kappa
\lambda})\nonumber.
\end{eqnarray}
In the gauge $\partial_0 A_0 = 0$ (which is satisfied by the Polyakov gauge) and
for temporally periodic spatial components of the gauge field $\vec A (\vec x,
T) = \vec A (\vec x, 0)$, the Pontrjagin index $\nu [A]$ (\ref{X11}) equals the
winding number of the Polyakov loop $\Omega (\bar x)$ \cite{4}:
\begin{eqnarray}
n [\Omega] &=& \frac{1}{24 \pi^2} \int d^3 x \epsilon_{ijk} \text{Tr} L_i L_j L_k\\
(L_i &=& \Omega \partial_i \Omega^\dagger)\nonumber.
\end{eqnarray}
For this winding number to be well defined, $\Omega(\vec x)$ has to
approach at spatial infinity $\vert \vec x \vert \to \infty$ a value independent
of the direction $\hat x$, so that the $3$-space ${\mathbbm{R}}^3$ can be topologically 
compactified to the 3-sphere $S_3$. Without loss of generality we can assume
that $\Omega (\vec x)$ approaches a center element
\begin{eqnarray}
\Omega (\vec x)\xrightarrow[|\vec x|\to\infty]{ } (-1)^{n_0}\label{X13}
\end{eqnarray}
with $n_0$ integer. In \cite{4} the following exact relation between the winding
number $n[\Omega]$ and the magnetic monopole charges has been derived
\begin{eqnarray}
n [\Omega ] = \sum_i l_i m_i.\label{X14}
\end{eqnarray}
Here the summation runs over all magnetic monopoles, $m_i$ being the
monopole charges, and the integer
\begin{eqnarray}
l_i = n_i - n_0
\end{eqnarray}
is defined by the center element $\Omega (\bar x_i ) = (-1)^{n_i}$ which the Polyakov
loop becomes at the monopole position $\bar x_i$ (cf. eq. \ref{X3}), and by the
boundary condition eq. \ref{X13}. For an isolated monopole with a Dirac string
running to infinity the quantity $l_i$ represents the length of the Dirac
string in group space.

To illustrate the above relation (\ref{X14}) consider the familiar instanton,
for which the temporal component of the gauge field is given by
\begin{eqnarray}
A_0^a (\bar x) = \frac{2x^a}{x_0^2 + r^2+ \rho^2}, \quad r = \vert \vec x \vert,
\end{eqnarray}
with $\rho$ being the instanton size. The Polyakov loop for the instanton 
field acquires the form of a hedgehog
\begin{eqnarray}
\Omega (\bar x) &=& \exp (i \chi (r) \hat x \vec \tau)\label{X17},\\
 \chi (r) &=& \pi
\frac{r}{\sqrt{r^2 + \rho ^2}}\nonumber
\end{eqnarray}
In the center of the hedgehog there is a magnetic monopole $\Omega (\bar x = 0)
= 1$, for which $n_{i=1} = 0$, and which carries a magnetic charge $m_{i=1} = 
1$. Furthermore, the hedgehog, eq. \ref{X17}, satisfies the boundary condition
eq. \ref{X13} with $n_0 = 1$. In this case the length of the Dirac string in
group space becomes $l=0-1=-1$ and the winding number becomes $n[\Omega]= -1$,
which is the correct result.

For the sake of illustration let us also consider a monopole-anti-monopole pair with
opposite magnetic charges connected by a Dirac string. We denote the value
of the Polyakov loop at the monopole and anti-monopole positions by
$\Omega = (-1)^n$ and $\Omega = (-1)^{\bar n}$, respectively. From eq. \ref{X14} we then
find for the winding number 
\begin{eqnarray}
n[\Omega] &=& (n-n_0) m-(\bar n - n_0) m\nonumber\\
 &=& (n - \bar n) m.
\end{eqnarray}
Again the winding number of the Polyakov loop field is given by the product of
the magnetic charge $m$ times the length of the Dirac string $n-\bar n$
connecting monopole and anti-monopole. The asymptotic value of the Polyakov loop
field at spatial infinity (\ref{X13}) drops out from this expression.

Finally let us replace our spatial manifold ${\mathbbm{R}}^3$ by the compact space
$S_3$. In this case the point at spatial infinity becomes part
of the manifold, and there is an extra monopole due to our boundary condition
at spatial infinity, eq.
\ref{X13}, collecting all the Dirac monopoles running to
infinity. Noticing that on a compact manifold the net charge has to vanish,
$\sum_{i=0} m_i = 0$, the monopole at infinity must carry the opposite net 
charge $m_0$ of all the remaining magnetic monopoles $m_0 = - \sum_{i \not= 0} 
m_i$. Eq. \ref{X14} then simplifies to 
\begin{eqnarray}
n [\Omega] = \sum_{i=0} n_i m_i,\label{X19}
\end{eqnarray}
where the summation over the monopoles now also includes the monopole at
infinity $(i=0)$.

Eq. \ref{X14} shows that in the Polyakov gauge the topology of gauge fields
is exclusively determined by the magnetic monopoles. Given the fact that all
Abelian gauges considered so far show the dual Meissner effect for the 
Yang--Mills vacuum we conjecture that this result is not restricted to the 
Polyakov gauge but applies to all Abelian gauges. The above considerations show
that magnetic monopoles are entirely sufficient to generate a non-trivial
topology and hence a non-zero topological susceptibility. Therefore, magnetic
monopoles are also sufficient to trigger spontaneous breaking of chiral symmetry,
which is usually attributed to instantons. Instantons in Abelian gauges give
rise to magnetic monopoles as we have seen above, but monopoles can exist in the
absence of instantons. The above considerations show that in the Abelian gauges
both confinement and spontaneous breaking of chiral symmetry can be generated by
the same field configurations: magnetic monopoles.
\nopagebreak
\begin{samepage}
\section{Acknowledgment}
Discussions with M. \mbox{Engelhardt}, K. Langfeld, M. Quandt, A. Sch\"afke and
O. \mbox{Tennert} are greatly acknowledged.

\end{samepage}
\end{document}